\documentclass[final,5p,times,twocolumn]{elsarticle}

\usepackage{graphicx}
\usepackage{graphicx,lscape}
\usepackage{amssymb}
\usepackage{url}
\usepackage{multirow}
\usepackage{scrextend}
\usepackage{lineno}
\usepackage{lscape}
\usepackage{kotex}
\usepackage{color}
\usepackage{verbatim}
\usepackage{listings}
\usepackage[htt]{hyphenat}
\usepackage{makecell}

\begin{document}

\begin{frontmatter}

\title{A Forensic Methodology for Detecting Image Manipulations}

\author{Jiwon Lee}
\author{Seungjae Jeon}
\author{Yunji Park}
\author{Jaehyun Chung}
\author{Doowon Jeong\corref{cor1}}
\address{College of Police and Criminal Justice, Dongguk University, Seoul, 04620, South Korea}

\cortext[cor1]{Corresponding author.\\
E-mail address: doowon@dgu.ac.kr (D. Jeong)}

\begin{abstract}
By applying artificial intelligence to image editing technology, it has become possible to generate high-quality images with minimal traces of manipulation. However, since these technologies can be misused for criminal activities such as dissemination of false information, destruction of evidence, and denial of facts, it is crucial to implement strong countermeasures. In this study, image file and mobile forensic artifacts analysis were conducted for detecting image manipulation. Image file analysis involves parsing the metadata of manipulated images (e.g., Exif, DQT, and Filename Signature) and comparing them with a Reference DB to detect manipulation. The Reference DB is a database that collects manipulation-related traces left in image metadata, which serves as a criterion for detecting image manipulation. In the mobile forensic artifacts analysis, packages related to image editing tools were extracted and analyzed to aid the detection of image manipulation. The proposed methodology overcomes the limitations of existing graphic feature-based analysis and combines with image processing techniques, providing the advantage of reducing false positives. The research results demonstrate the significant role of such methodology in digital forensic investigation and analysis. Additionally, We provide the code for parsing image metadata and the Reference DB along with the dataset of manipulated images, aiming to contribute to related research.
\end{abstract}


\end{frontmatter}

\section{Introduction}
\label{sec:introduction}
The proliferation of digital technology has brought about a surge in the accessibility of image editing tools, enabling individuals to easily edit visual content. Notable applications such as Adobe Photoshop, Snapseed, and Meitu have become readily available for use. Furthermore, the use of smartphones has rendered basic image editing tools commonplace, without requiring the download of additional third-party apps from app stores. These applications offer a variety of editing functions including cropping, resizing, filters, mosaic, color correction, and brightness correction.

In recent times, the application of AI technology in image editing has gained significant popularity, in addition to the aforementioned editing methods. A prevalent technique that utilizes this technology is known as inpainting, which involves `object removal' and `background removal'. These tasks are primarily performed using deep learning algorithms \cite{qin2021image}. The principle of object removal is to identify the pixel values corresponding to the object to be removed from the image and replace them with the pixel values surrounding the area to be removed. In this replacement process, the relationships with adjacent pixel values are taken into consideration to generate a natural-looking image. Similarly, the background removal technology operates on the same principle, with the task of distinguishing between objects and backgrounds performed using deep learning algorithms, followed by the removal of the recognized background parts. 

\begin{figure}[h] 
    \centering
    \includegraphics[width=\linewidth]{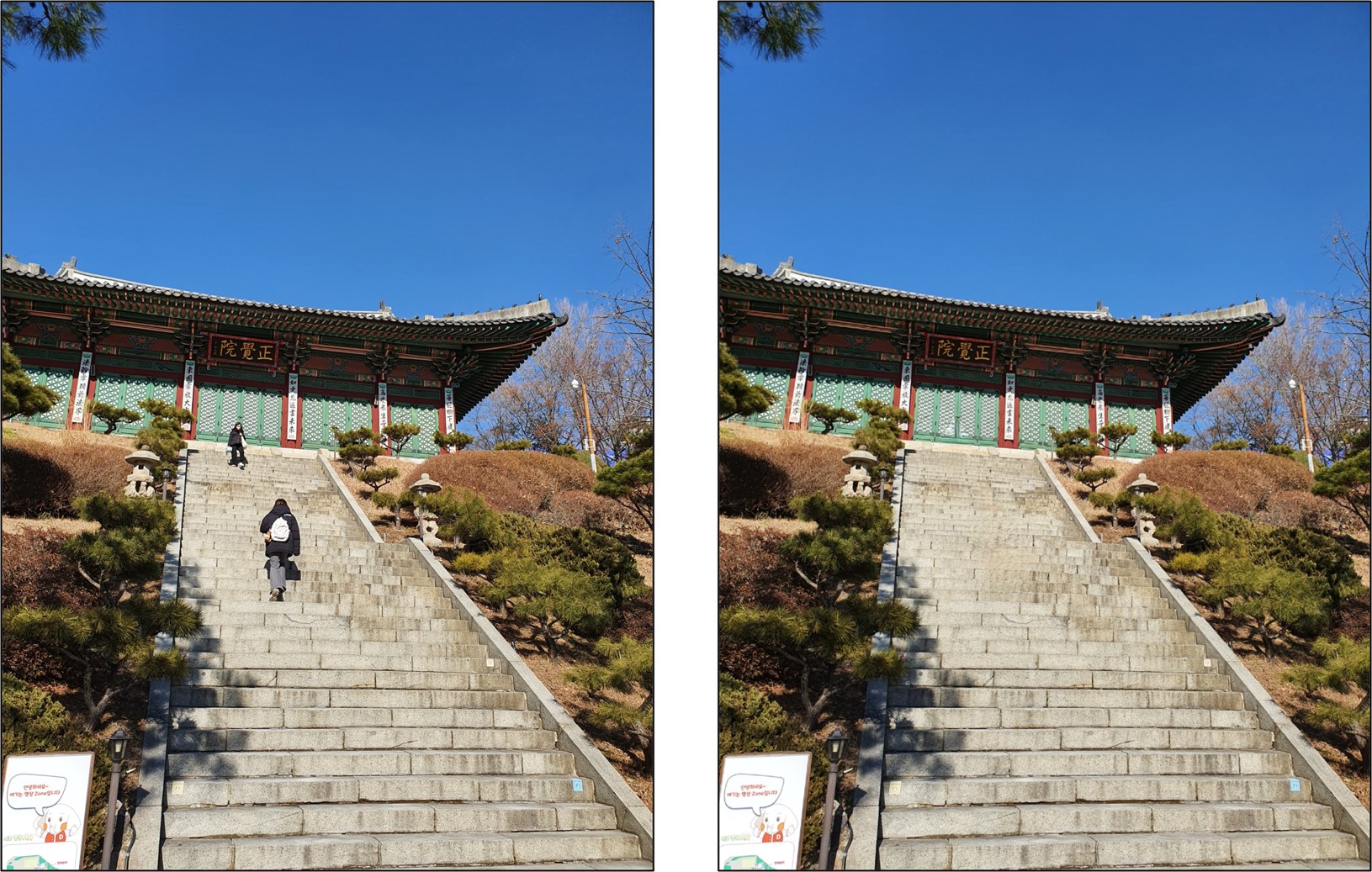}
    \caption{Original image (Left) and edited image with the person object removed (Right)}
    \label{fig:editing_example1}
\end{figure}

\begin{figure}
    \centering
    \includegraphics[width=\linewidth]{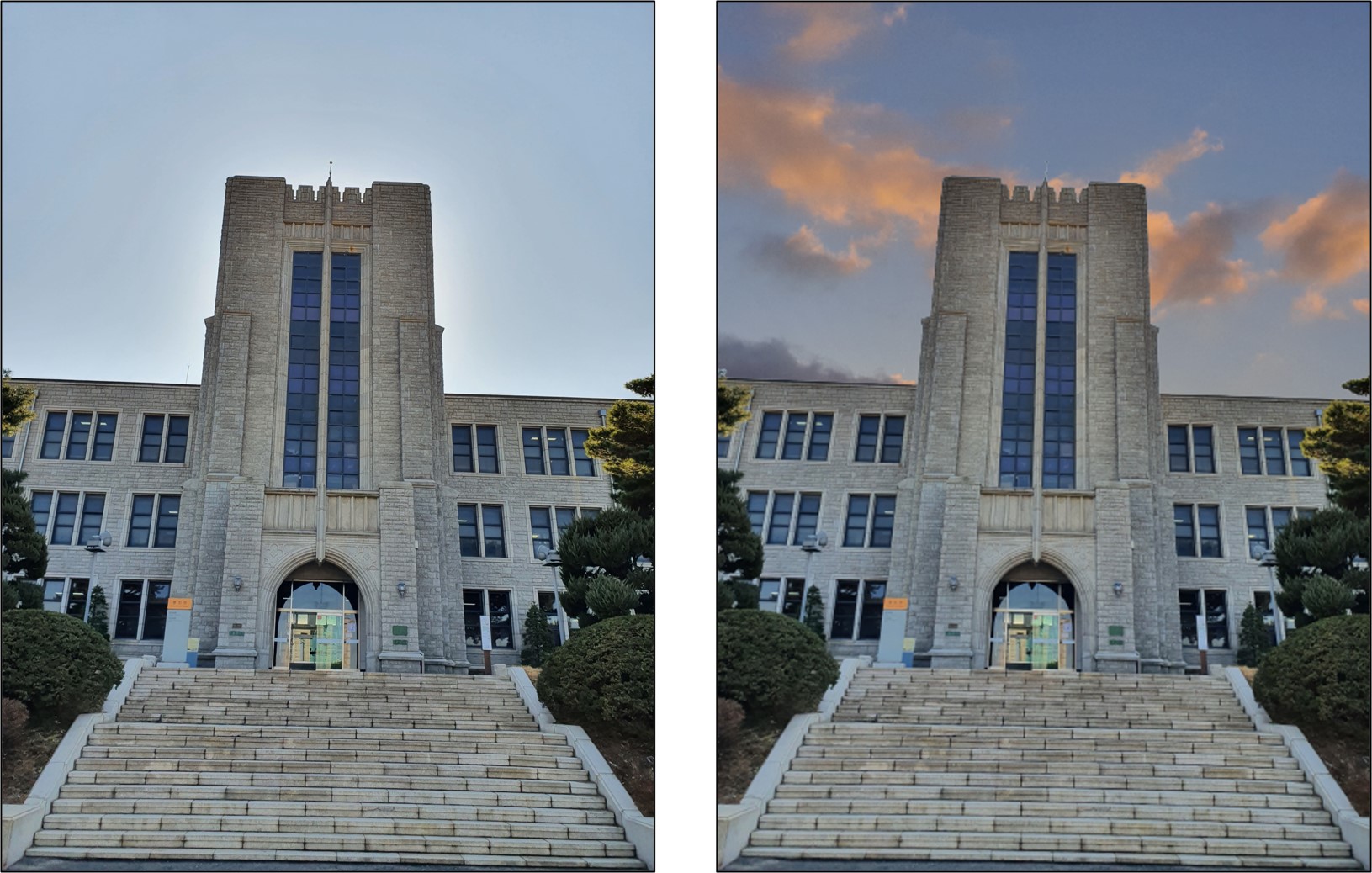}
    \caption{Original image (Left) and edited image where the original background has been removed and replaced with another one (Right)}
    \label{fig:editing_example2}
\end{figure}

Figure \ref{fig:editing_example1} and \ref{fig:editing_example2} depict images that have undergone editing using image editing tools. Figure \ref{fig:editing_example1} shows an image in which the object `person' has been removed using the object removal function, while Figure \ref{fig:editing_example2} shows an image in which the background has been removed and replaced with a different background using the background removal function. When exclusively examining the edited image, it is discernible that it has been thoroughly removed and altered to an extent where the presence of editing becomes indiscernible. Previously, in order to enhance the quality of edited images, it was necessary to manually edit individual pixels by specifying snippet areas. However, currently, anyone can easily achieve high-quality results, with edited elements such as sharpness and color representation exhibiting little to no noticeable impact on the edited image. 

On the other hand, high accessibility to these tools means that anyone can manipulate images, which implies the potential misuse for criminal activities such as the dissemination of false information, destruction of evidence, and denial of facts. Therefore, it is necessary to implement robust countermeasures. In this context, image manipulation detection techniques are gaining considerable attention in the field of image forensics \cite{zheng2019survey}. However, as these techniques primarily focuses on the graphic analysis of the image itself, it is limited in its ability to provide a comprehensive methodology for detecting image manipulation within the context of digital forensics. In this paper, we aim to proactively respond to related criminal investigations by analyzing mobile forensic artifacts in addition to image file analysis for detecting image manipulation.

The remainder of the paper is organized as follows. Section \ref{sec:background} presents previous studies on image manipulation techniques and detection methods. Section \ref{sec:imagefileanalysis} describes an approach for creating a dataset of manipulated images and explains how manipulation can be detected through image file analysis. Additionally, Section \ref{sec:mobileforensic} presents an approach for detecting manipulation by analyzing mobile forensic artifacts. Section \ref{sec:process} introduces a methodology for detecting image manipulation that is proposed in this study and conducts a discussion on it. Finally, Section \ref{sec:conclusion} provides a summary of the key points of this paper and suggests directions for future research.
\section{Background and motivation}
\label{sec:background}
In this section, we examine the previous studies on image manipulation and present the motivation of this paper by comparing it with those studies.

\subsection{Image manipulation}\label{subsec:manipulation}
\begin{figure} 
    \centering
    \includegraphics[width=\linewidth]{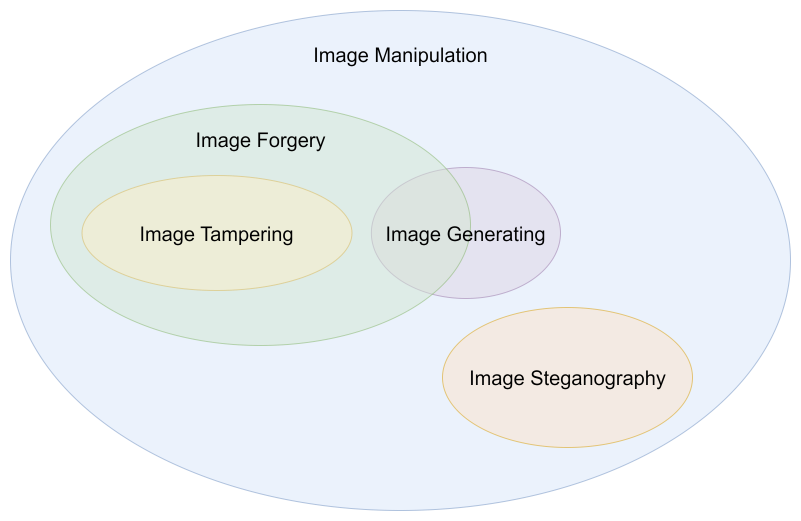}
    \caption{Image manipulation category}
    \label{fig:manipulation_category}
\end{figure}

Image manipulation refers to any actions taken on a digital image using an image editing tool on a digital device. Figure \ref{fig:manipulation_category} shows the subcategories of image manipulation. Zheng et al. \cite{zheng2019survey} and Thakur and Rohilla \cite{thakur2020recent} have classified image manipulation into subcategories, such as image forgery, image tampering, image generating, and image steganography. Image forgery refers to image manipulation that creates false graphical content to deceive people about past events. Image tampering is a specialized form of image forgery that involves altering specific parts of an image's graphics. Image generating is a technique in which an image is computer-generated, or parts of an image are computer-generated and can be used to forge an image. Image steganography refers to the technique of embedding additional information in an image by changing or filling certain pixels in an image with specific values, rather than using graphical characteristics to deceive the human eye like image forgery. 

Commonly used techniques for image forgery include copy-move, cut-paste, and erase-fill \cite{zheng2019survey,thakur2020recent}. Copy-move involves copying a particular object or region of an image and pasting it somewhere else in the same image. This technique can be used to add or remove objects from an image or to create multiple copies of an object within the same image. Cut-paste is another technique that is frequently used in image forgery, involving the cutting and pasting of a specific object or region of an image into another image. This technique can be used to create a composite image or to add or remove objects from an image. Erase-fill is a technique that is used to alter an image by erasing certain objects or regions of an image and filling them in to match their surroundings. This technique aims to maintain the consistency of the image after the object has been removed, without making it stand out awkwardly. Both copy-move and cut-paste techniques are encompassed by the term `Splicing', while erase-fill is referred to as `Inpainting' \cite{alahmadi2013splicing,guillemot2013image}.

Numerous studies have been conducted on detecting image forgery, with two commonly used methods being the keypoint-based approach and the block-based approach. The keypoint-based approach involves extracting feature points from an image and detecting any potential forgery through matching these points. This approach is particularly useful for post-processing tasks such as image resizing and rotation and allows for the detection of manipulated regions when compared to the original image. Feature extraction algorithms such as SIFT and SURF are often used in conjunction with techniques for matching feature points, including nearest-neighbor and clustering \cite{abd2016copy, amerini2011sift, yang2018copy, huang2008detection, bo2010image, shivakumar2011detection}. The block-based approach, on the other hand, involves dividing an image into pixel blocks of a certain size, extracting the features of each block, and comparing the similarity between blocks to detect any potential forgery. Techniques for extracting feature points of pixel blocks include frequency domain, dimensionality reduction, texture, moment invariant, and image local features. Techniques for finding similarity between feature points include sorting, hashing, and Euclidean distance \cite{zheng2019survey, thakur2020recent, gupta2013detecting, huang2011improved, cao2012robust, soni2018cmfd, li2007sorted}. Unlike the copy-move forgery detection, cut-paste forgery detection detects traces left by the manipulation process since there are no identical duplicate regions in one image. Typical traces include abnormal edges, inconsistent lighting, JPEG compression, and traces of the camera used to capture the image \cite{zheng2019survey}. To detect erase-fill forgery, it is possible to identify the blur that remains when a certain region is removed or the erasure of another region by duplicating it, using the aforementioned block-based approach \cite{zheng2019survey}.

There are also studies that focus not on the graphic features of images but rather on the characteristics inherent in the image file format. Particularly, research has been conducted on detecting the source of capturing devices and applications through the image generation process of the widely used JPEG file format in smartphones. Kim et al. \cite{kim2016building} conducted a comprehensive analysis of the DQT generated in the process of JPEG image creation, both in the image itself and in the image thumbnail. They confirmed the ability to identify the device used for image capture and the application used for storage. Hur et al. \cite{Hur2020jpeg} analyzed the JPEG compression algorithm to study the processing process of MMS and messenger images based on the device that disseminated the image. This analysis allowed them to identify the specific application and device manufacturer used for image dissemination. Previous studies have primarily focused on image analysis, mainly utilizing metadata or compression algorithms to identify the source of the image. However, this approach only emphasizes the information recorded within the image file, thereby failing to present a comprehensive methodology for detecting image manipulation.

In this paper, we prioritize an approach that requires relatively fewer resources for algorithm development and execution compared to deep learning-based methodologies. We present a comprehensive forensic methodology for image manipulation detection, taking into consideration the specificity of digital forensics.

\subsection{Motivation}\label{subsec:motivation}
In Figure \ref{fig:manipulation_category}, it can be observed that the term `image manipulation' is used to encompass all actions that alter digital images. While the subject of digital forensic analysis can occasionally be the image itself, it primarily focuses on the storage media. Within such storage media, numerous manipulated images may exist, and in order to detect forged images, it is necessary to first detect and select manipulated images. Furthermore, as confirmed in Section \ref{subsec:manipulation}, image forgery refers to the creation of fake graphical content. However, since not all actions performed on an image can be considered solely for the purpose of generating fake graphical content, this paper addresses a broader scope of image manipulation detection rather than solely focusing on image forgery detection.

In recent times, many studies have been conducted on the detection of image manipulation. However, most of previous studies have concentrated on detecting image manipulation when given access to the original image. These studies have analyzed the graphical characteristics of the image to detect signs of manipulation. In real-world investigations, however, it may be necessary to analyze the manipulated image itself without access to the original image. Furthermore, if the image has been subjected to numerous manipulations, it may be challenging to identify abnormal graphics regions, making it difficult to confirm if the image has been manipulated. Despite these challenges, image file analysis can provide an effective means of detecting image manipulation, even if the original image is not available, graphical detection is not possible, and the detection algorithm is unknown. The approach described in this study does not rely on deep learning for abnormality detection, but instead utilizes the metadata of image files. Consequently, this methodology offers advantages in terms of computational efficiency and processing speed when compared to methods that utilize deep learning models. Such research is not only useful for investigators but also for the general public.

In the field of digital forensics, various scenarios and situations related to image manipulation need to be considered. For instance, in the case of digital media such as images or videos used as evidence in court, if the authenticity of the images is in doubt, it is necessary to analyze mobile forensic artifacts to determine when and how the images were manipulated. Furthermore, when dealing with illegal adult content distributed online, it may be necessary to seize the perpetrator's smartphone and analyze mobile forensic artifacts to determine whether the images have been manipulated or not. In such situations, relying solely on image file analysis may not provide sufficient information, and therefore, it is necessary to analyze artifacts from third-party apps to detect which images were used and how they were manipulated by the image forger. 

Therefore, this study aims to conduct exactly the detection of image manipulation by conducting analysis of image file and mobile forensic artifacts. By doing so, it seeks to contribute to the advancement of image manipulation detection techniques in the field of digital forensics and their application in criminal investigations and analysis.
\section{Image file analysis for image manipulation detection}
\label{sec:imagefileanalysis}
This section presents an approach for detecting image manipulation through analysis of image files. In this study, we create a manipulated image dataset and analyze the metadata of each image to detect traces of manipulation. In addition, we create a reference database based on the analysis results, which serves as a criterion for image manipulation.

\subsection{Manipulated image dataset}\label{subsec:dataset}
We conducted a study on the Android operating system, which holds the largest global market share among smartphone platforms. The dataset was created by directly editing images using 11 distinct image editing tools. Among these tools, 10 applications used in the study had a rating of 4.0 or higher and garnered over 1 million downloads on the Google Play Store. Additionally, we utilized a photo editor supported by Samsung Galaxy. Information regarding the image editing tools employed in this study is provided in Table \ref{tab:software_info}. Out of the 11 applications, 7 utilized the object removal function, while the remaining 4 applications employed the background removal function to produce manipulated images.

\begin{table*}
\resizebox{\textwidth}{!}{
\begin{tabular}{lllll}
\hline
APP & Corp. & Version & Package & Function \\ \hline
Snapseed                    & Google LLC              & 2.19.1.303051424 & com.niksoftware.snapseed               & Object Removal       \\ 
Meitu                       & Meitu(China) Limited    & 9.7.5.5          & com.mt.mtxx.mtxx                       & Object Removal       \\ 
Remove Unwanted Object      & BG.Studio               & 1.3.8            & vn.remove.photo.content                & Object Removal       \\ 
SnapEdit                    & SnapEdit Team           & 3.4.1            & snapedit.app.remove                    & Object Removal       \\ 
Adobe Photoshop Fix         & Adobe                   & 1.1.0            & com.adobe.adobephotoshopfix            & Object Removal       \\ 
Photoshop Express           & Adobe                   & 8.8.17           & com.adobe.psmobile                     & Object Removal       \\ 
Samsung Photo Editor        & Samsung                 & 3.0.25.33        & com.sec.android.mimage.photoretouching & Object Removal       \\ 
removebg                    & Kaleido AI              & 1.1.4            & bg.remove.android                      & Background Removal   \\ 
Background Eraser (Inshot)   & InShot                  & 2.122.33         & photoeditor.cutout.backgrounderaser    & Background Removal   \\ 
Background Eraser (handy)    & handy Closet            & 4.1.0            & com.handycloset.android.eraser         & Background Removal   \\ 
Photo Studio                & KVADGroup App Studio   & 2.6.2.1178        & com.kvadgroup.photostudio              & Background Removal   \\ 
\hline
\end{tabular}
}
\caption{Image editing tools information}
\label{tab:software_info}
\end{table*}

When generating manipulated images for the dataset, we paid close attention to the post-processing functions and saving methods offered by the image editing tools during the image saving process. This is because metadata and configuration values can differ depending on the image settings. Table \ref{tab:postprocessing_method} displays the setting values that were considered for each image editing tool and the total number of manipulated image samples that were generated.

In the process of generating manipulated images for this study, 11 original images were captured using a Galaxy S10e smartphone. Furthermore, the manipulated images were generated by considering the post-processing functions and saving methods provided by each image editing tool, as outlined in Table \ref{tab:postprocessing_method}. The number of sample images generated through each image editing tool was determined using the formula: \texttt{Number of original images} $\cdot$ \texttt{Number of setting values} $\cdot$ \texttt{Number of save methods}. In total, a dataset comprising 968 images was created.

\begin{figure} 
    \centering
    \includegraphics[width=\linewidth]{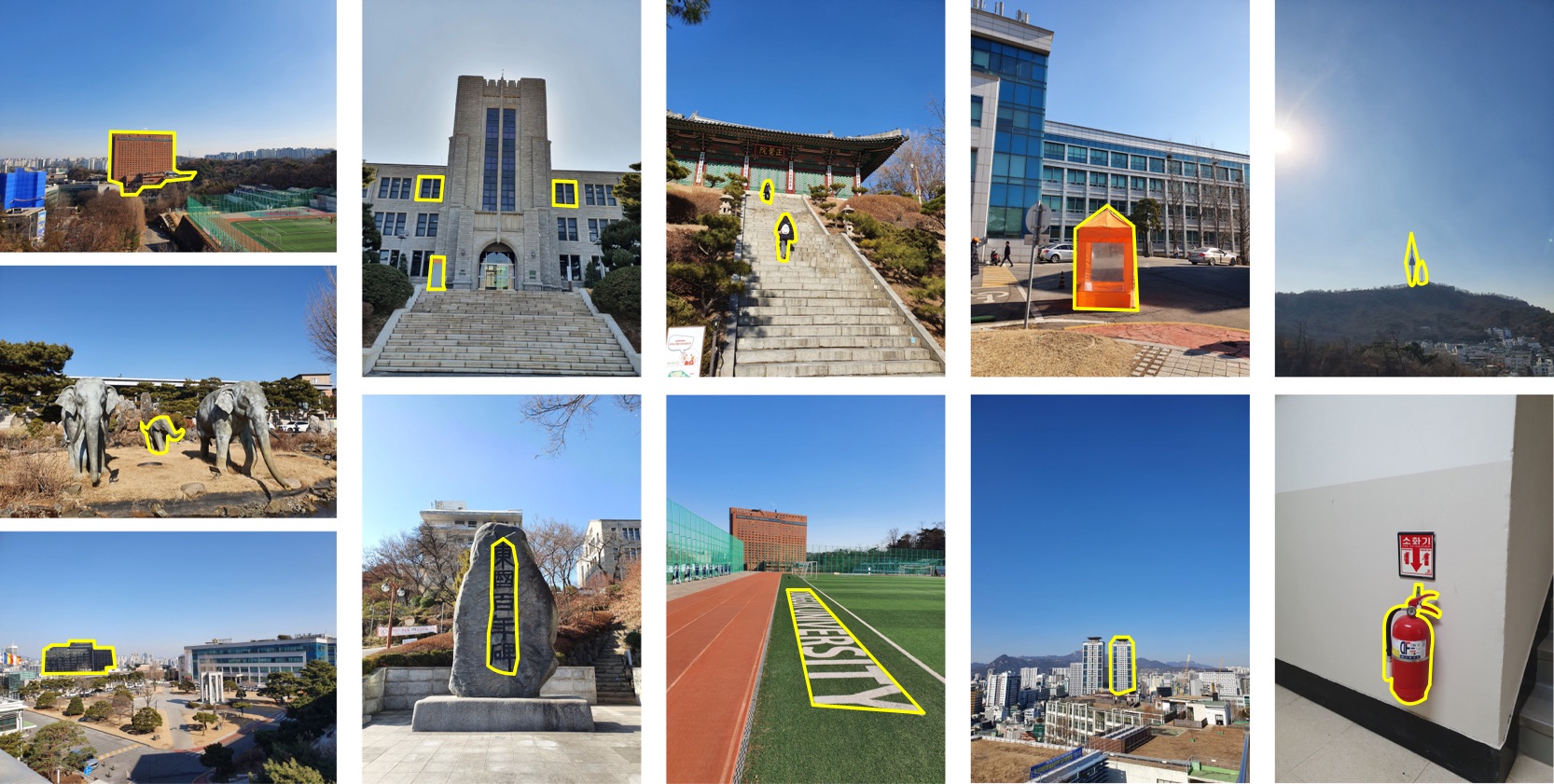} 
    \caption{Segmentation of objects}
    \label{fig:seg_object}
\end{figure}

\begin{figure} 
    \centering
    \includegraphics[width=\linewidth]{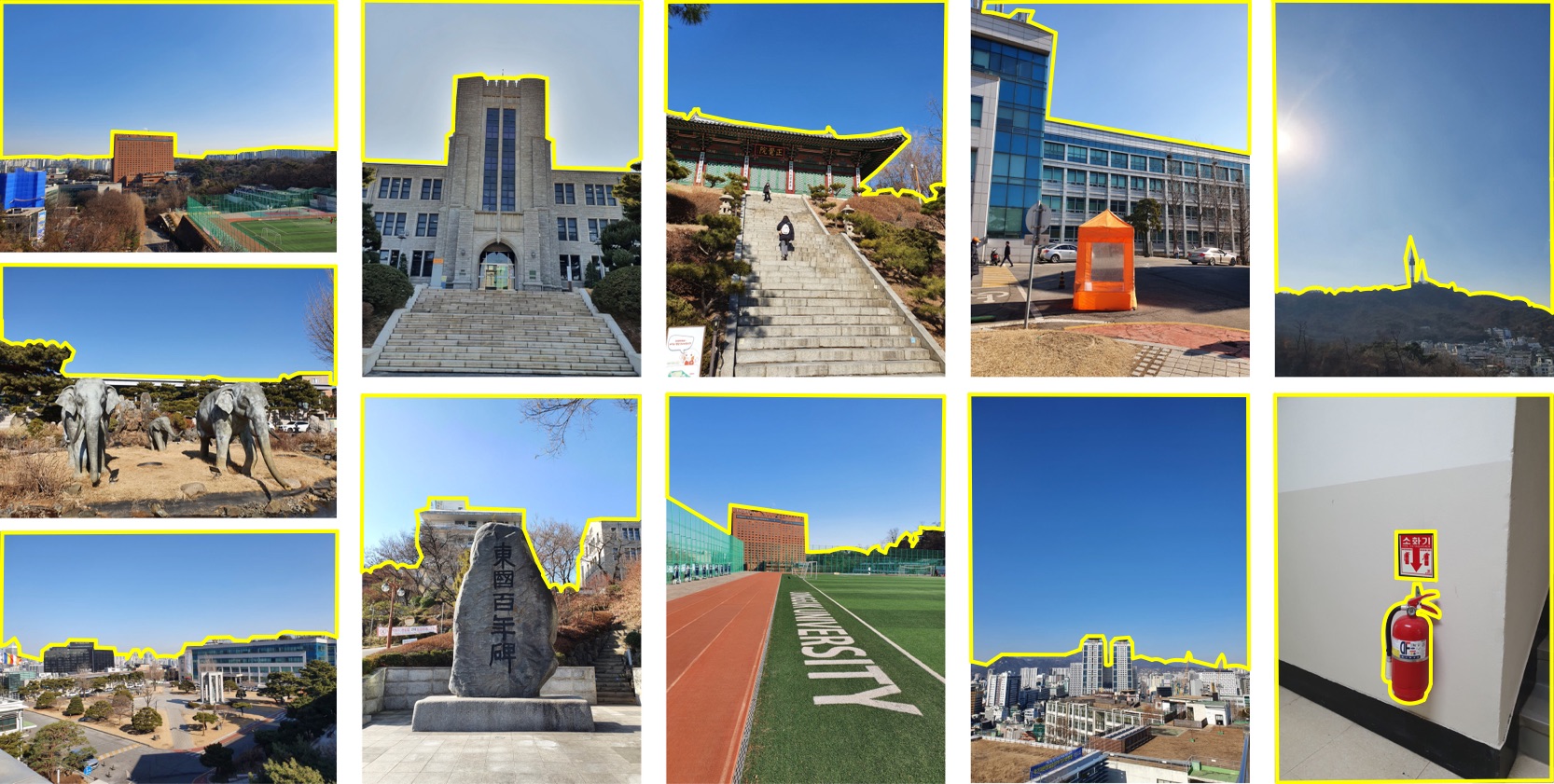}
    \caption{Segmentation of backgrounds}
    \label{fig:seg_background}
\end{figure}

Figure \ref{fig:seg_object} exhibits the segmented object that is intended to be removed from 11 original images used in generating the manipulated image, while Figure \ref{fig:seg_background} shows the segmented background that serves as the target of the removal.

The manipulated image dataset created in this study and files containing information on how to save each image can be downloaded from the GitHub repository\footnote{\url{https://github.com/allinonee/Manipulated-Image-Dataset}}.

\begin{table*}
\resizebox{\textwidth}{!}{
\begin{tabular}{llll}
\hline
APP & Setting Values & Save Methods & Number of samples \\ 
\hline
Snapssed                  & \begin{tabular}[t]{@{}l@{}}Image Size (800, 1366, 1920, 2000, 4000, No Resizing)\\ Format \& Quality (JPG 100\%, 95\%, 80\%, PNG)\end{tabular} & \begin{tabular}[t]{@{}l@{}}Save, Export,\\ Export to another folder\end{tabular} & 539 \\
Meitu                     & Image Quality (UHD, Standard) & Save, Quick Save  & 44  \\
Remove Unwanted Object    & · & Save & 11  \\
SnapEdit                  & · & High, Standard & 22  \\
Adobe Photoshop Fix       & · & Save to Gallery & 11  \\
Photoshop Express         & \begin{tabular}[t]{@{}l@{}}Image Size (600x800, 1125x1500, 1500x2000, 300x4000,\\ No Resizing, Square)\end{tabular} & Save & 66  \\
Samsung Photo Editor      & Image Size (20\%, 40\%, 60\%, 80\%, No Resizing) & Save, Save as another file & 11  \\
Background Eraser (Inshot) & Image Size (1080, 1920) & Save & 22  \\
Background Eraser (handy)  & Smooth Edge (0, 1, 2, 3, 4, 5) & Save & 66  \\
Photo Studio              & Image Size (Normal, Small, No Resizing) & JPG, PNG & 66  \\
removebg                  & · & Download (preview image) & 110 \\
\hline
\end{tabular}
}
\caption{Post-processing functions and save methods considered for generating manipulated images}
\label{tab:postprocessing_method}
\end{table*}

\subsection{Image file analysis}\label{subsec:imagefile}
Utilizing the dataset created in Section \ref{subsec:dataset}, we analyze the traces of manipulation present in the metadata of image files resulting from image manipulation. This study focuses on conducting a study on the Joint Photographic Experts Group (JPEG), which is widely recognized as the standard file format in digital image processing. JPEG includes not only the scanned data of an image but also metadata containing information about the corresponding image. Therefore, in this section, we explain the fundamental concept of the file format structure as well as the metadata that can be utilized to detect image manipulation.

\subsubsection{Define Quantization Table (DQT)}\label{subsubsec:dqt}
We employ Define Quantization Table (DQT) for the purpose of detecting image manipulation. We present an introduction to the JPEG compression process and file format structure in order to facilitate a better understanding of DQT.

\begin{figure} 
    \centering
    \includegraphics[width=\linewidth]{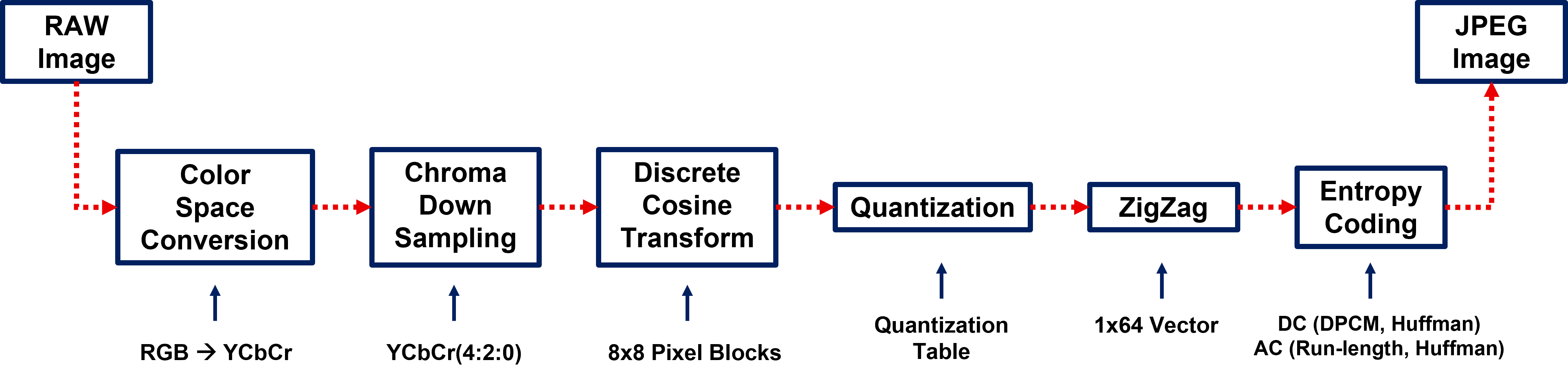}
    \caption{JPEG compression process}
    \label{fig:JPEGprocess}
\end{figure}

Figure \ref{fig:JPEGprocess} provides a visual representation of the JPEG compression process. Initially, JPEG converts the color space of an image from the RGB color model to the YCbCr (or YUV) color model. The image then undergoes chroma down sampling, a process that reduces color data to enhance compression efficiency. During this process, the Y component, representing brightness information, remains unaltered, while the Cb and Cr components, representing color information, are reduced according to a specific ratio. Subsequently, the image is partitioned into smaller blocks through Discrete Cosine Transform (DCT), allowing for the extraction of its frequency components. These components are then subjected to quantization to discard insignificant information. Following this, the data undergoes ZigZag Scanning, which transforms the quantized data into a one-dimensional sequence. Ultimately, the compressed data is generated using the Huffman encoding algorithm. The resulting JPEG image possesses an internal structure, as depicted in Table \ref{tab:JPEGformat} \cite{gloe2012forensic}.

DQT is a quantization table within the JPEG file format that determines the compression rate of an image. It is represented as an 8x8 integer array and is used to quantize the coefficients of pixel blocks obtained from performing DCT. DQT typically consists of two tables for luminance and chrominance, which can be utilized as unique digital identifiers derived from camera models and manufacturers \cite{kornblum2008using,kim2016building,kim2018mobile}. These characteristics can also be applied to identify image editing tools. Since image editing tools utilize specific DQT to determine image quality, they can serve as reference points for identifying the source and manipulation of an image. Table \ref{tab:example_DQT} presents the analysis results of the manipulated image dataset, showcasing representative DQT provided by different image editing tools. It was observed that image editing tools either used the same DQT or employed their own unique DQT. Such patterns can be effectively utilized for detecting image manipulation. For instance, if a particular DQT pattern appears in multiple images, it increases the likelihood that those images have been manipulated, thereby enhancing the accuracy of image manipulation detection.

\begin{table}
\resizebox{\columnwidth}{!}{
\begin{tabular}{lll}
\hline
Name & Bytes & Full Name \\ 
\hline
SOI         & 0xF FD8  & Start of Image                     \\
APP\#       & \begin{tabular}[t]{@{}l@{}}0xFF E0 $\sim$ \\ 0xFF E15\end{tabular} & App0 (JFIF), App1 (Exif)    \\
DQT         & 0xFF DB  & Define Quantization Table(s)       \\
SOF0        & 0xFF C0  & Start of Frame (baseline DCT)      \\
SOF         & 0xFF C2  & Start of Frame (progressive DCT)   \\
D           & 0xFF C4  & Define Huffman Table(s)            \\
SOS         & 0xFF DA  & Start of Scan                      \\
Scan Data   & \multicolumn{1}{c}{-} & Image Data            \\
DRI         & 0xFF DD  & Define Restart Interval          \\
RST\#       & \begin{tabular}[t]{@{}l@{}}0xFF D0 $\sim$ \\ 0xFF D7\end{tabular}  & Restart     \\
COM         & 0xFF FE  & Comment                            \\
EOI         & 0xFF D9  & End of Image                       \\
\hline
\end{tabular}
}
\caption{JPEG file format}
\label{tab:JPEGformat}
\end{table}

\begin{table*}
\resizebox{\textwidth}{!}{%
\begin{tabular}{c|c|c}
\hline
App & \begin{tabular}[t]{@{}l@{}}Snapseed, Meitu, Adobe Photoshop Fix, \\ Background Eraser (Inshot), Photo Studio\end{tabular} & \begin{tabular}[t]{@{}l@{}}Remove Unwanted Object, \\ SnapEdit, Photo Studio\end{tabular} \\
\hline
DQT & & \\
& \includegraphics[width=0.4\textwidth]{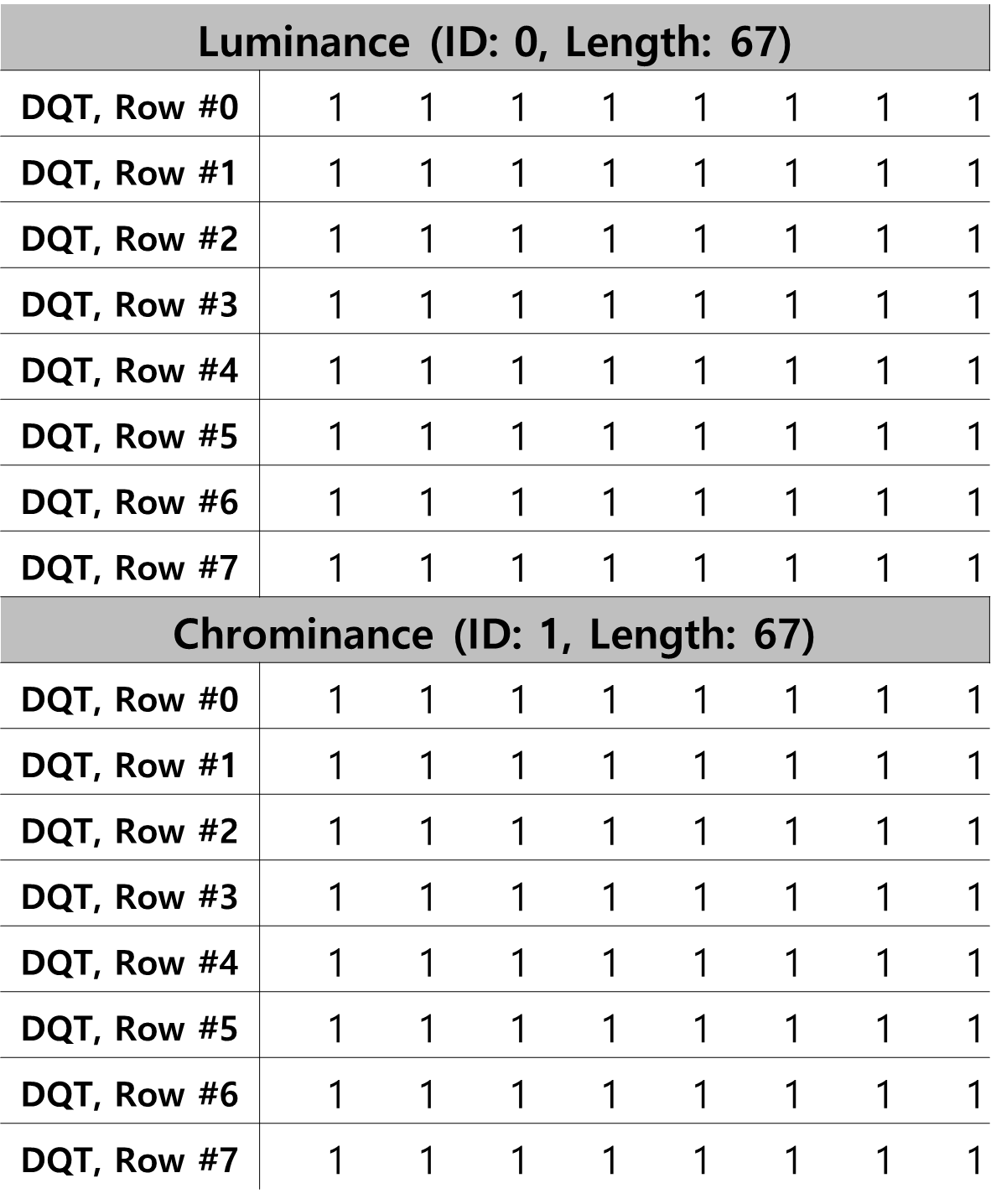} & 
\includegraphics[width=0.4\textwidth]{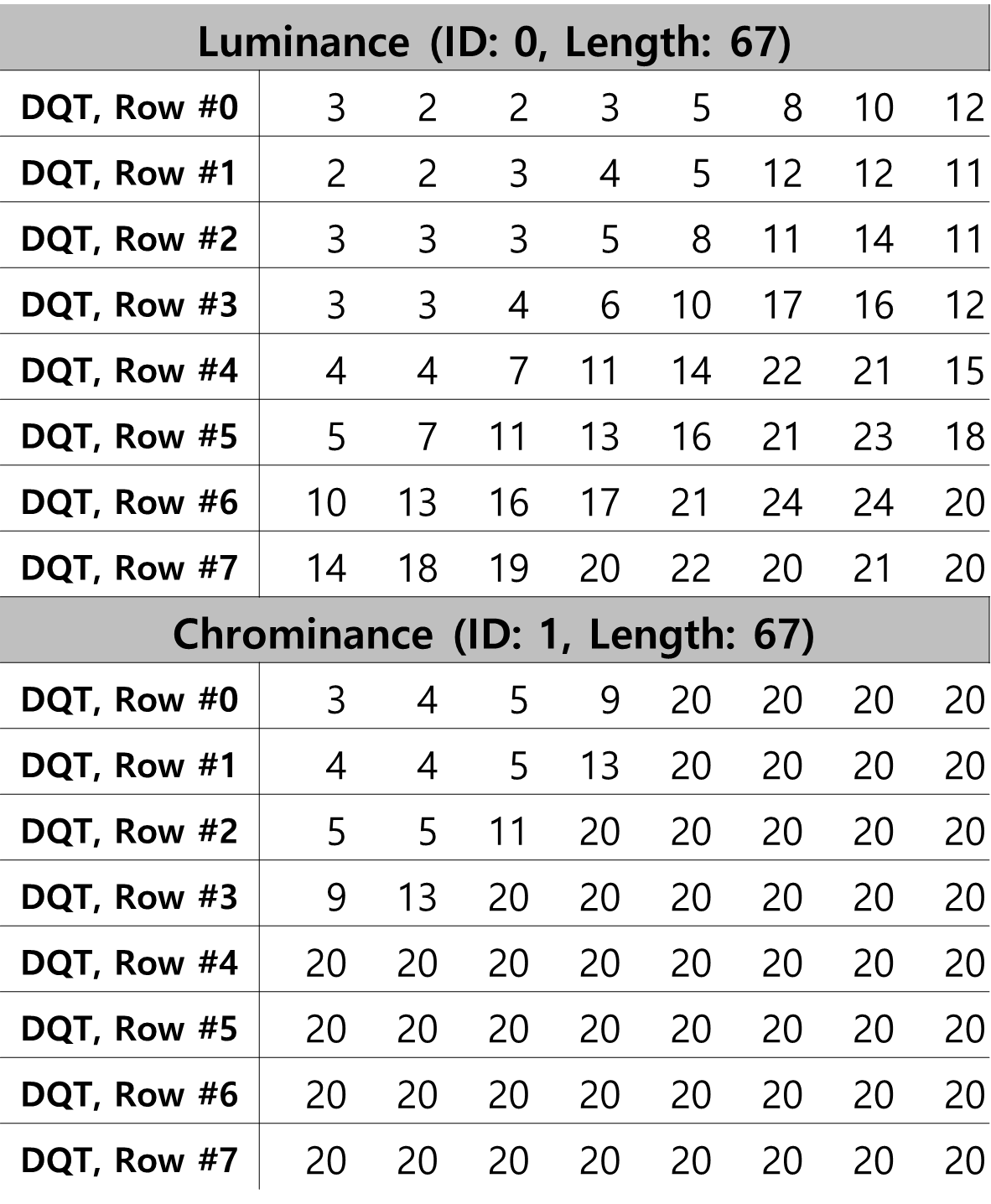} 
\\                     
\hline
\end{tabular}
}
\caption{Representative DQT used by image editing tools}
\label{tab:example_DQT}
\end{table*}

\subsubsection{Exchangeable Image File Format (Exif)}\label{subsubsec:exif} 
We employ the Exchangeable Image File Format (Exif) for the purpose of detecting image manipulation. Exif is a metadata standard that provides additional information about digital images, including details about the device used to take the image. When an image is edited using an image editing tool, the tool's signature is often left behind in the Exif, which enables the detection of image manipulation. Information on image editing tools that leave a signature in the Exif is presented in Table \ref{tab:Exifsig}. Out of the 11 image editing tools examined in this study, we found that 4 applications included a signature that could be traced back to the tool. Although Exif is commonly used to identify the source or editing history of an image, it is susceptible to human manipulation.

\begin{table}
\resizebox{\columnwidth}{!}{
\begin{tabular}{ll}
\hline
App & Exif \\ 
\hline
Snapseed & Software : Snapseed 2.0 \\
Meitu & \begin{tabular}[t]{@{}l@{}}Artist : Meitu\\ Software : Meitu 9755\end{tabular} \\
Remove Unwanted Object & Software : AdvaSoft TouchRetouch  \\
Photoshop Express      & Software : Adobe Photoshop Express (Android) \\
\hline
\end{tabular}
}
\caption{Exif containing signatures}
\label{tab:Exifsig}
\end{table}

\subsubsection{Filename signature}\label{subsubsec:filenamesignature}
This study examined the filenames of manipulated images to not only identify whether the images were manipulated but also to determine which image editing tool was utilized. The outcomes of analyzing the filenames of the images edited by each image editing tool are presented in Table \ref{tab:editedimage_filename}. The findings reveal that the filenames of the edited images mainly include the date of creation, the original image filename used for editing, and the signature of the image editing tool. Although it is possible for individuals to modify filenames, which can limit the forensic analysis, there are circumstances where filenames are preserved, and the unique filenames created by editing tools can provide valuable clues.

\begin{table*}
\resizebox{\textwidth}{!}{
\begin{tabular}{lllll}
\hline
APP & Save Option & Edited Image Filename Info & Filename Signature \\ 
\hline
Snapseed & \begin{tabular}[t]{@{}l@{}} Save \\ Export to another folder\end{tabular} & \begin{tabular}[t]{@{}l@{}} (Original\_image\_filename)-(Number).jpeg \\ (Original\_image\_filename)\_edited.(jpeg or png) \end{tabular} & \begin{tabular}[t]{@{}l@{}} . \\ edited \end{tabular} \\

Meitu & \begin{tabular}[t]{@{}l@{}} Save \\ Quick Save\end{tabular} & \begin{tabular}[t]{@{}l@{}} MTXX\_MH(Edited\_image\_creation\_datetime).jpg \\ MTXX\_formula(Edited\_image\_creation\_datetime).jpg \end{tabular} & \begin{tabular}[t]{@{}l@{}} MTXX, MH, \\ formula \end{tabular} \\

Remove Unwanted Object      & Save & \begin{tabular}[t]{@{}l@{}}WipeOut(Edited\_image\_creation\_minute)\_(Edited\_image\_creation\_day)\_\\ (Edited\_image\_creation\_year)\_(Edited\_image\_creation\_time).jpg\end{tabular} & WipeOut    \\

SnapEdit                    & Save  & (Edited\_image\_creation\_datetime).png  & .   \\

Adobe Photoshop Fix         & Save  & PSFix\_(Edited\_image\_creation\_date)\_(Edited\_image\_creation\_time).jpeg  & PSFix    \\

Photoshop Express           & Save  & PSX\_(Edited\_image\_creation\_date)\_(Edited\_image\_creation\_time).jpg     & PSX      \\

removebg                    & Download  & ei\_(Edited\_image\_creation\_datetime)-removebg-preview.png  & removebg    \\

Background Eraser (Inshot)   & Save  & \begin{tabular}[t]{@{}l@{}}BackgroundEraser\_(Edited\_image\_creation\_date)\_\\ (Edited\_image\_creation\_time).jpg\end{tabular} & BackgroundEraser  \\

Background Eraser (handy)    & Save  & (Edited\_image\_creation\_datetime).png  & .     \\

Photo Studio                & Save  & photostudio\_(Edited\_image\_creation\_datetime).(jpg   or png)  & photostudio     \\

Samsung Photo Editor & \begin{tabular}[t]{@{}l@{}} Save \\ Save as another file\end{tabular} & \begin{tabular}[t]{@{}l@{}} (Original\_image\_creation\_date)\_(Original\_image\_creation\_time).jpg \\ (Edited\_image\_creation\_date)\_(Edited\_image\_creation\_time).jpg \end{tabular} & \begin{tabular}[t]{@{}l@{}} . \\ . \end{tabular} \\

\hline
\end{tabular}
}
\caption{Edited images filename analysis}
\label{tab:editedimage_filename}
\end{table*}

\subsection{Reference Database}\label{subsec:referenceDB}
Taking into comprehensive consideration the metadata of image files, we aim to detect image manipulations and determine the ultimate source of the image. Therefore, we developed DQT Parser in this study which can parse the Exif and DQT of manipulated images, subsequently analyzed the parsed results to insert data into the Reference DB. Additionally, in cases where filenames contain image editing tool signatures, we also input the information. To insert manipulation-related data into the initial Reference DB, the manipulated image dataset created in Section \ref{subsec:dataset} is utilized. However, considering the potential changes in Exif, DQT, and Filename Signature due to software updates, regular DB updates are conducted to ensure accurate detection of manipulated images. The Reference DB thus created serves as the criterion for detecting image manipulations and determining the ultimate source of the image.

\begin{figure} 
    \centering
    \includegraphics[width=\linewidth]{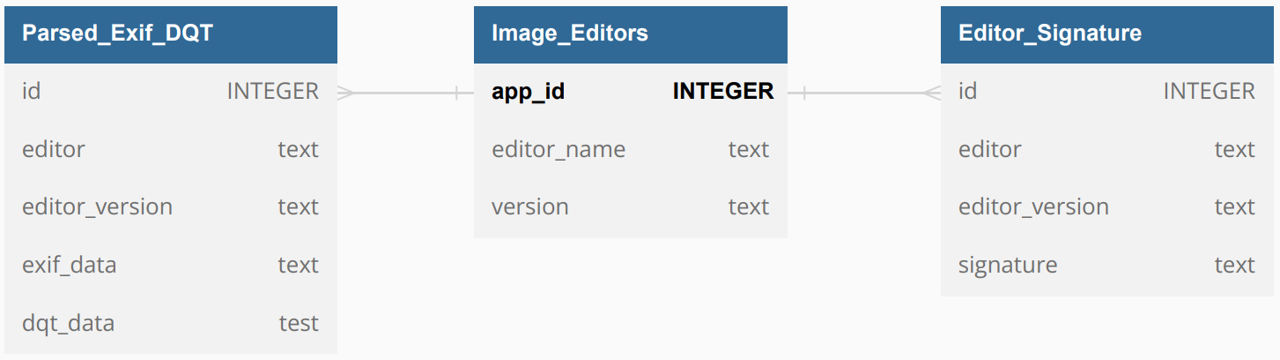}
    \caption{Schema for the Reference DB}
    \label{fig:schema}
\end{figure}

Figure \ref{fig:schema} shows the schema for the Reference DB. The Reference DB includes tables such as Image\_Editors, Parsed\_Exif\_DQT, and Editor\_Signature. The Image\_Editors table serves as a reference table designed to store the names and corresponding versions of various image editing tools. The Parsed\_Exif\_DQT table is responsible for storing the Exif and DQT data specific to each image editing tool. During the Exif data processing, the information indicative of the image editing tool's signature was parsed using the Exiftool utility. As for the DQT data, direct parsing was performed by examining the DQT header signature within the JPEG image, followed by the application of a hash function (MD5) to generate and insert the resulting hash value. In the case of the Editor\_Signature table, relevant information is inserted if the signature of an image editing tool is present in the image filename.

Detailed information on the DQT Parser developed in this study and Reference DB created can be found in the GitHub repository\footnote{https://github.com/allinonee/DQT-Parser}.
\section{Mobile forensic artifacts analysis for image manipulation detection}
\label{sec:mobileforensic}
In this section, we present an approach for detecting image manipulation through mobile forensic artifacts analysis. 

\subsection{Considerations}\label{subsec:considerations}
To gather comprehensive information on image manipulation, this study focuses on analyzing the package path where app data of image editing tools is stored. The path `/data/data/\{package\_name\}/' in Android stores data of apps with corresponding package names, including settings, logs, cache, and databases. After editing images with 11 image editing tools utilized in Section \ref{sec:imagefileanalysis}, this study extracted and analyzed the path and conducted additional analysis if another path contained significant data.

In this paper, the following information has been considered to detect image manipulation. The first consideration is the existence of an edited image. In this context, the term `edited image' refers not to the images stored in the gallery, but rather to the artifacts left by the image editing tool itself, which are present in the path where the app data of the image editing tool is stored. When an image is edited, it is automatically saved to the gallery. However, if the image is deleted from the gallery, the edited image cannot be identified. Therefore, the existence of an edited image even after the deletion of the image stored in the gallary can be a significant piece of information to respond to anti-forensics activities such as evidence destruction. The second consideration is the identification of the manipulated region. By examining the manipulated image dataset created in this study, it is apparent that determining whether an image has been manipulated with can be challenging if the original image is not available. Thus, being able to detect the exact regions of the image that have been manipulated can be very beneficial for investigators when digital images are used as evidence. The existence of an original image is also considered. By identifying the image used for editing, investigators can compare the similarity of the edited image to the original image, allowing them to determine whether the edited image was derived from the original image.

\begin{table*}
\resizebox{\textwidth}{!}{
\begin{tabular}{lcccccccc}
\hline
APP & Edited image & Manipulated region & Original image & Edited logs & Image caching & Account info & Installation time & Recent usage time \\ 
\hline
Snapseed                 & $\triangle$   & . & . & . & . & . & . & . \\
Meitu                    & O & . & O & O & O & . & O & O  \\
Remove Unwanted Object   & . & . & . & . & O & . & O & O  \\
SnapEdit                 & O & O & O & . & . & . & O & O  \\
Adobe Photoshop Fix      & . & . & O & O & O & O & O & O  \\
Photoshop Express        & . & . & . & . & O & O & O & O  \\
Samsung Photo Editor     & . & . & O & . & . & . & . & .  \\
removebg                 & . & . & O & . & . & . & O & O  \\
Background Eraser (Inshot) & $\triangle$  & . & $\triangle$  & O & O & . & O & O  \\
Background Eraser (handy)  & . & . & $\triangle$  & . & . & . & O & O   \\
Photo   Studio           & O & O & O & O & O & . & O & O  \\
\hline
\end{tabular}
}
\caption{Information that can be confirmed by image editing tool (The triangle shape indicates that only the most recently edited record can be checked)}
\label{tab:software_artifacts_info}
\end{table*}

\begin{figure*} 
    \centering
    \includegraphics[width=\linewidth]{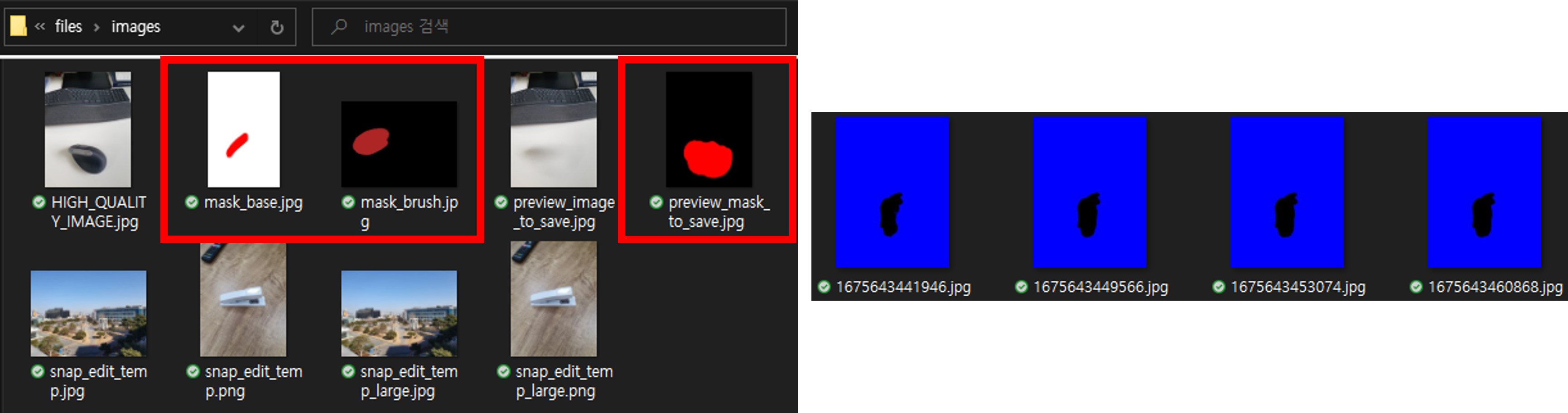}
    \caption{Identification of manipulated regions through mask}
    \label{fig:case_mask}
\end{figure*}

The fourth consideration is the existence of edit logs. It is crucial to know when and how an image was edited and what the edited result was. Edit logs can be particularly helpful in artifacts analysis, providing concrete information about image manipulation. The image caching status is also considered. Most applications store cached images to enhance performance, decrease server load, and avoid capacity issues. This feature can be useful in detecting anti-forensics behavior. Image editing tools require permission to access the gallery during the first use to save the edited image to the gallery. Therefore, the cache includes not only images used by the app itself but also images stored in the gallery. Even if an image is deleted from the gallery to destruct evidence, cached images left behind by the app can be viewed to determine which images were in the gallery. In addition, we examined account information, installation time, and recent app usage time to aid in detecting manipulation.  

\subsection{Mobile forensic artifacts analysis}\label{subsec:Mobile forensic artifacts analysis}
We present the results of actual analysis based on the considerations presented. The analysis is performed for the 11 applications utilized in Section \ref{sec:imagefileanalysis}. Table \ref{tab:software_artifacts_info} presents the outcomes of our comprehensive analysis of mobile forensic artifacts related to image editing tools. 

\subsubsection{Edited images}\label{subsubsec:editedimages}
In 5 out of the 11 applications, we were able to detect the edited images. For instance, in the case of Snapseed, the edited image file was generated with a random number, and was detected by changing the extension through JPEG signature identification. Concerning Meitu and Background Eraser (Inshot), the files were found in the path `/storage/emulated/0/Android/data/\{package\_name\}', which is the private external storage. Furthermore, in the case of SnapEdit and BackgroundEraser (Inshot), only the most recently edited image could be ascertained, while Photo Studio left the edited image when editing an image with a project that saves the settings and work for editing.

\subsubsection{Manipulated regions}\label{subsubsec:editedareas}
Most of the applications did not leave any artifacts related to the manipulated region, but it was found that SnapEdit leaves a mask for the region where the object was removed, as shown in the left photo in Figure~\ref{fig:case_mask}, and Photo Studio also saves a mask for the most recently edited image as `\{EditTimeInfo\}.jpg', as shown in the right photo in Figure~\ref{fig:case_mask}. This can be helpful in identifying the specific actions taken by the user, such as whether they removed or retained certain objects in the image.

\begin{figure}[h] 
    \centering
    \includegraphics[width=\linewidth]{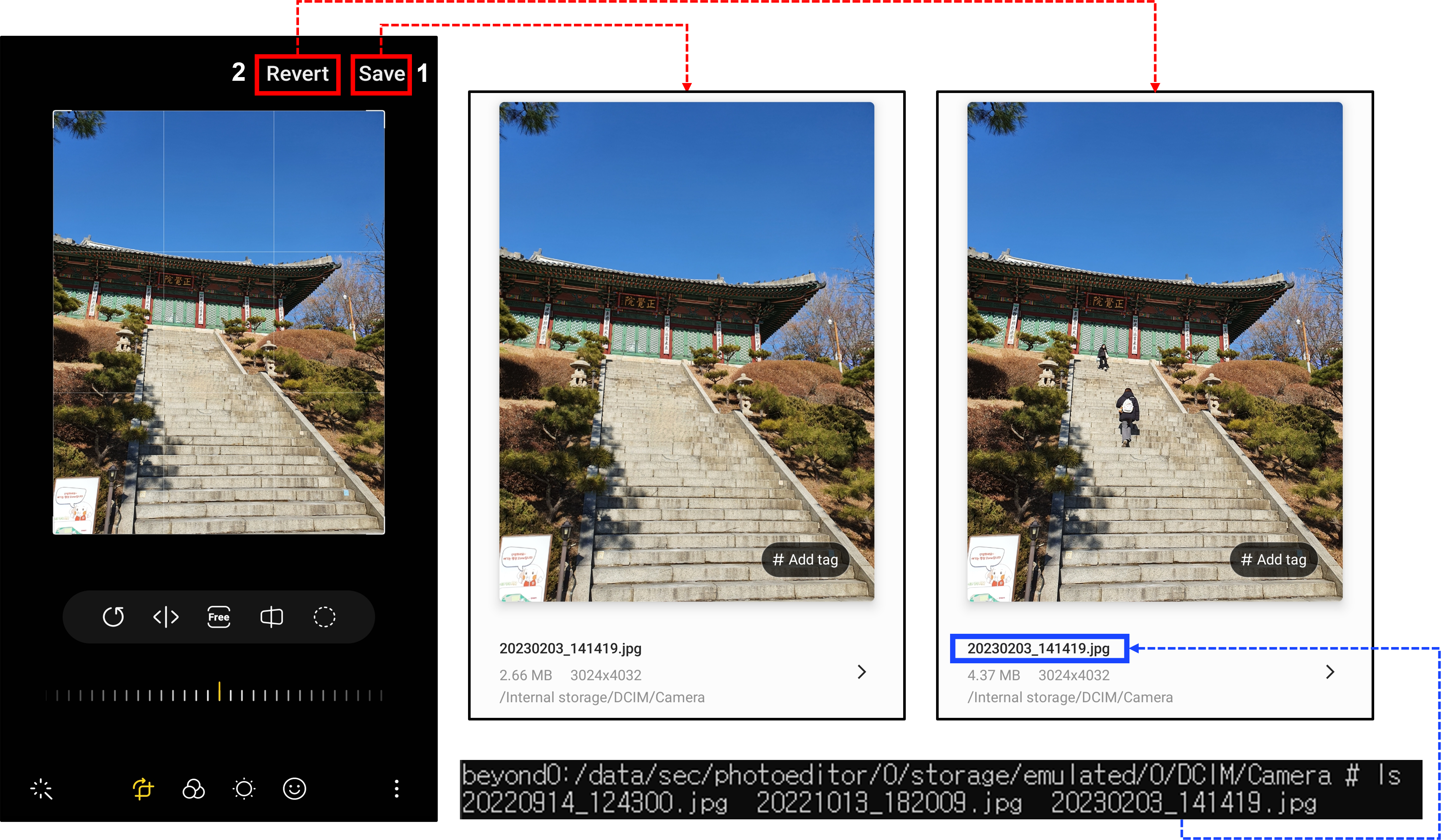}
    \caption{Original image identification path when using `Save’ function}
    \label{fig:case_samsungpath}
\end{figure}

\begin{figure*}[h] 
    \centering
    \includegraphics[width=\linewidth]{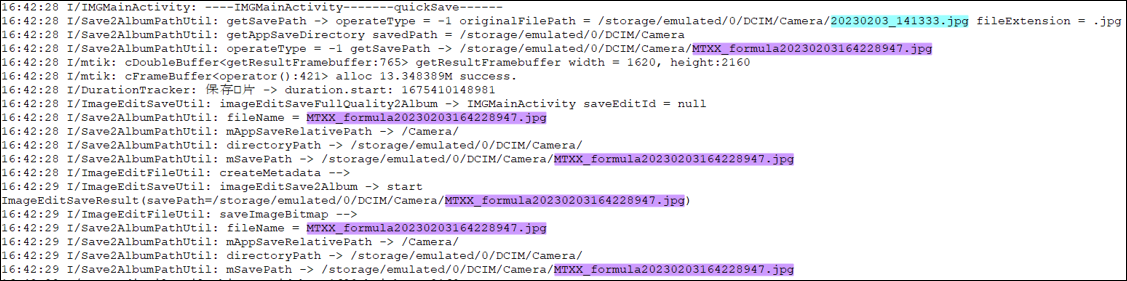}
    \caption{Edit logs recorded with timestamp}
    \label{fig:case_logfile}
\end{figure*}

\subsubsection{Original images}\label{subsubsec:originalimages}
In 8 out of the 11 applications, we were able to detect the original image used for editing. In most cases, the original image existed in the path where packages are stored. However, for Meitu and Background Eraser (Inshot), the original images were detected in the private external storage path. Samsung Photo Editor was also able to detect the original image from a specific path. This editor supports two saving methods: `Save', which overwrites the edited settings in the original file itself, and `Save as copy'. Generally, using the `Save' function allows users to restore the original image before editing through the `Revert' function, as shown in Figure \ref{fig:case_samsungpath}. We can also see that the original image is saved in `/data/sec/photoeditor/0/storage/emulated/0/DCIM/Camera/\{original\_image\_filename\}', as shown in the path in Figure \ref{fig:case_samsungpath}. Samsung Photo Editor is one of the basic image editing tools on Samsung smartphones, enabling users to perform high-quality image editing without downloading third-party apps. As the basic app is used by many users, analyzing these apps can significantly contribute to forensic investigations.

\subsubsection{Edit logs}\label{subsubsec:editlogs}
The role of edit logs in detecting image manipulation is crucial. For instance, the Meitu records information such as the storage path of the edited image, the filenames of the original and edited images, the start time of editing, and the editing functions used. Logs are also created in Adobe Photoshop Fix and Photo Studio when project-based editing is used, including the project name, creation time, and modification time. In addition, Background Eraser (Inshot) records information such as the edited image filename, editing start time, and save time. As these log files are saved with timestamps, as shown in Figure \ref{fig:case_logfile}, they can greatly aid in detecting image manipulation.

\begin{figure}[h]
    \centering
    \includegraphics[width=\linewidth]{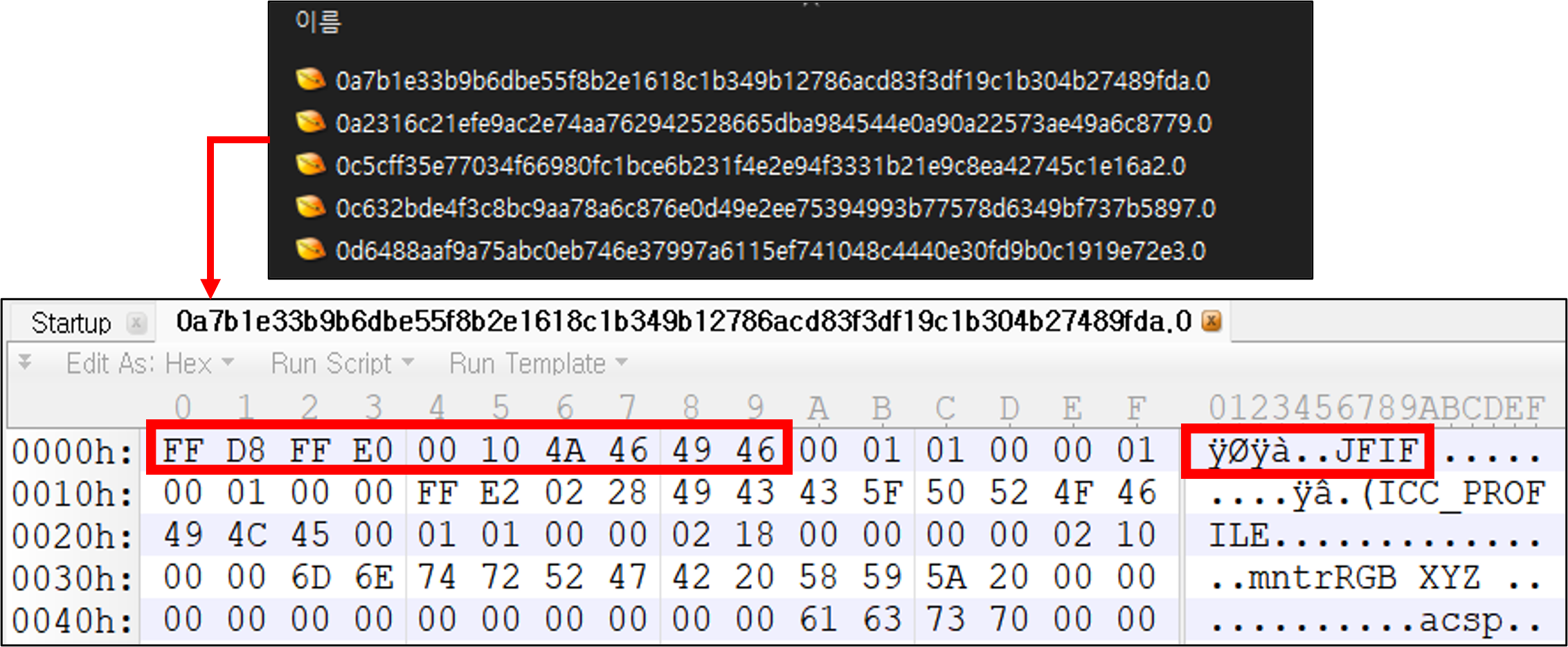}
    \caption{Cache file with JPEG signature}
    \label{fig:case_cachesig}
\end{figure}

\begin{figure} 
    \centering
    \includegraphics[width=\linewidth]{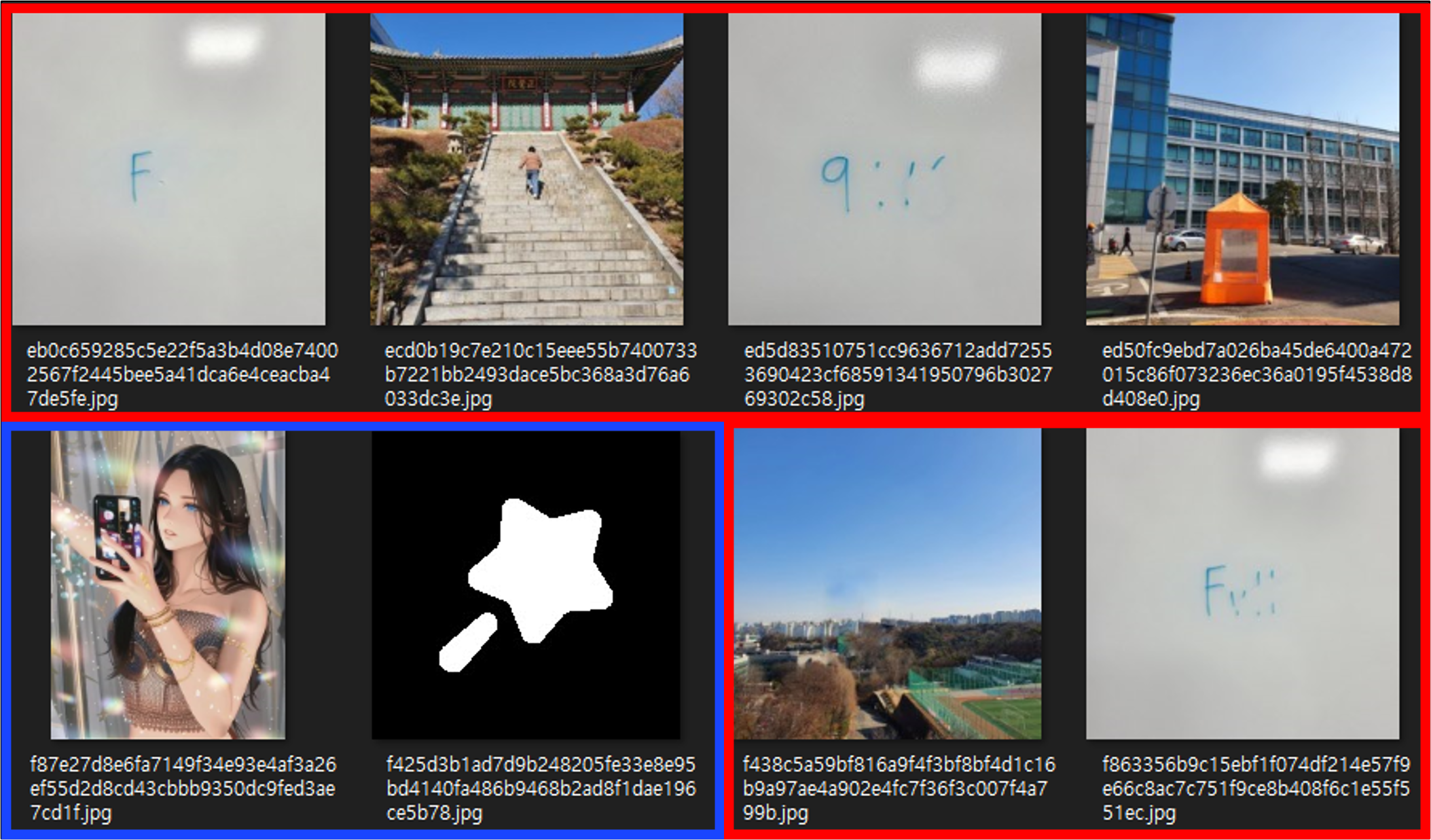}
    \caption{Image editing tool that even caches gallery images (red borders)}
    \label{fig:case_cachebord}
\end{figure}

\subsubsection{Image caching status}\label{subsubsec:caching}
Cached images can be useful in cases where the user has engaged in anti-forensic behavior. Typically, image editing applications request permission to access the gallery during their initial use to allow for the saving of edited images to the gallery. At this point, the images stored in the gallery are also cached, and even if the user deletes an image from the gallery, the cached images may still be accessible. The files in the path `/data/data/\{package\_name\}/cache/image\_manager\_disk\_cache/' where the cached images are stored are illustrated in Figure \ref{fig:case_cachesig}. All files possess the filename `\{random\_num\}.0' and feature a JPEG signature. If an attempt is made to view the images by changing the extension, both images used by the application itself (blue border) and those stored in the gallery (red border) may be visible, as shown in Figure \ref{fig:case_cachebord}.

\section{Methodology}
\label{sec:process} 
This section presents a comprehensive methodology for detecting image manipulation that includes the findings from Sections \ref{sec:imagefileanalysis} and \ref{sec:mobileforensic}. 

\subsection{Proposed methodology}\label{subsec:methodology}

\begin{figure*}[h]
    \centering
    \includegraphics[width=\linewidth]{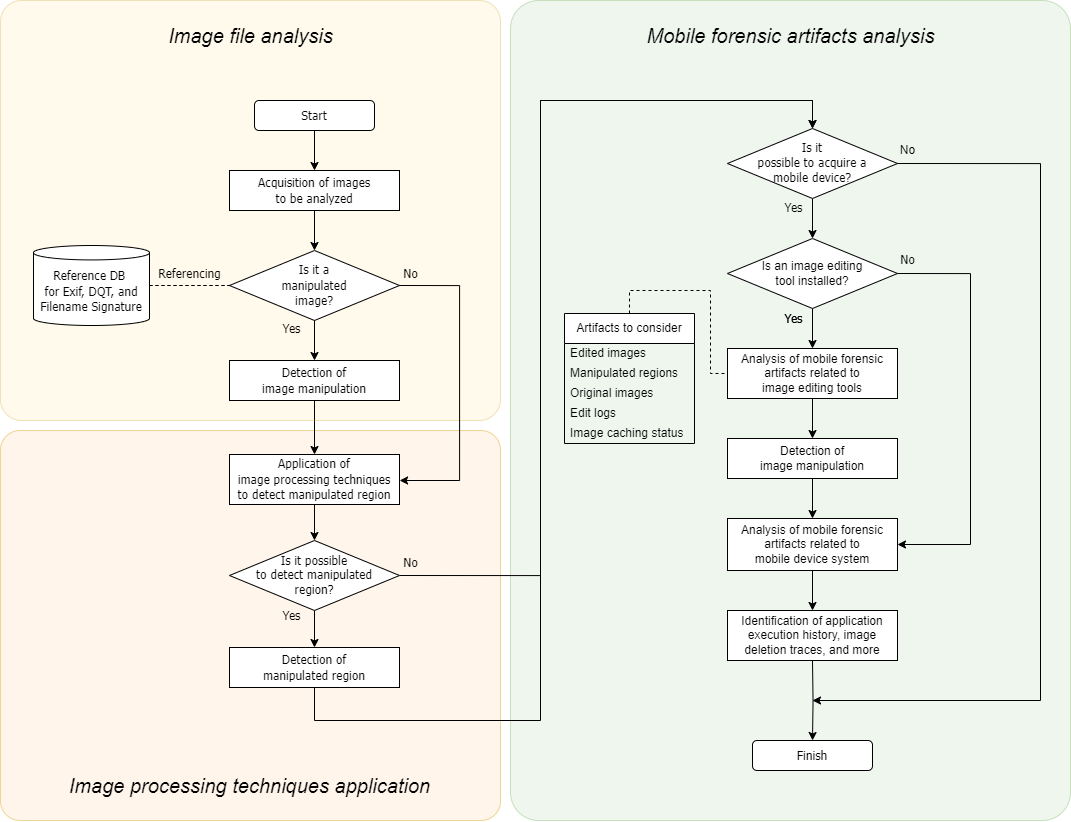}
    \caption{Methodology for image manipulation detection}
    \label{fig:detection_process}
\end{figure*}

Figure \ref{fig:detection_process} illustrates the proposed methodology for image manipulation detection in this study. The methodology consists of three stages, among which the image processing techniques application refers to the utilization of existing techniques that are employed for image manipulation detection.

The first stage, image file analysis, the target image is acquired, and to determine if the image has been manipulated, a reference database for manipulation detection is consulted. To accomplish this, the Exif, DQT, and Filename Signature of the target image are parsed and analyzed to detect the presence of manipulation. This process allows us to determine whether the image has been manipulated and provides information about the last known source of the image, including the image editing tool used.

The second stage, the application of image processing techniques, conventional image processing methods are employed to detect manipulated regions. Existing image processing techniques used for manipulation detection include Noise Analysis, Error Level Analysis (ELA), Principal Component Analysis (PCA), and Luminance Gradient. Through these techniques, abnormal graphic regions can be identified, enabling the detection of regions that have been altered due to manipulation.

The final stage, mobile forensic artifacts analysis, involves the detection of image manipulation by extracting and analyzing data from mobile devices with installed image editing tools. If a mobile device is obtained, the analysis focuses on considerations for detecting image manipulation, such as edited images, manipulated regions, original images, edit logs, and image caching status. These details can be analyzed from the extracted artifacts of the image editing tool. Additionally, analyzing system artifacts of the mobile device allows the identification of behaviors related to tool execution and image deletion, among others.

\subsection{Discussion}\label{subsec:discussion}
In this section, we examine the advantages and disadvantages of each technique presented in the methodology and discuss the need for complementary approaches among these techniques. Image file analysis and the application of image processing techniques involve analyzing image files, while mobile digital forensic artifact analysis involves analyzing mobile devices.

Image file analysis has the advantage of being simple and fast, based on the metadata signature within the file \cite{vadrevu2022image, abdanalysis}. However, it requires prior analysis for new applications, and in cases where file transfers or re-encoding with changes in image quality occur, the metadata can be compromised. Additionally, it has the disadvantage of being easily manipulated by users \cite{mani2022survey}.

\begin{figure}[h]
    \centering
    \includegraphics[width=\linewidth]{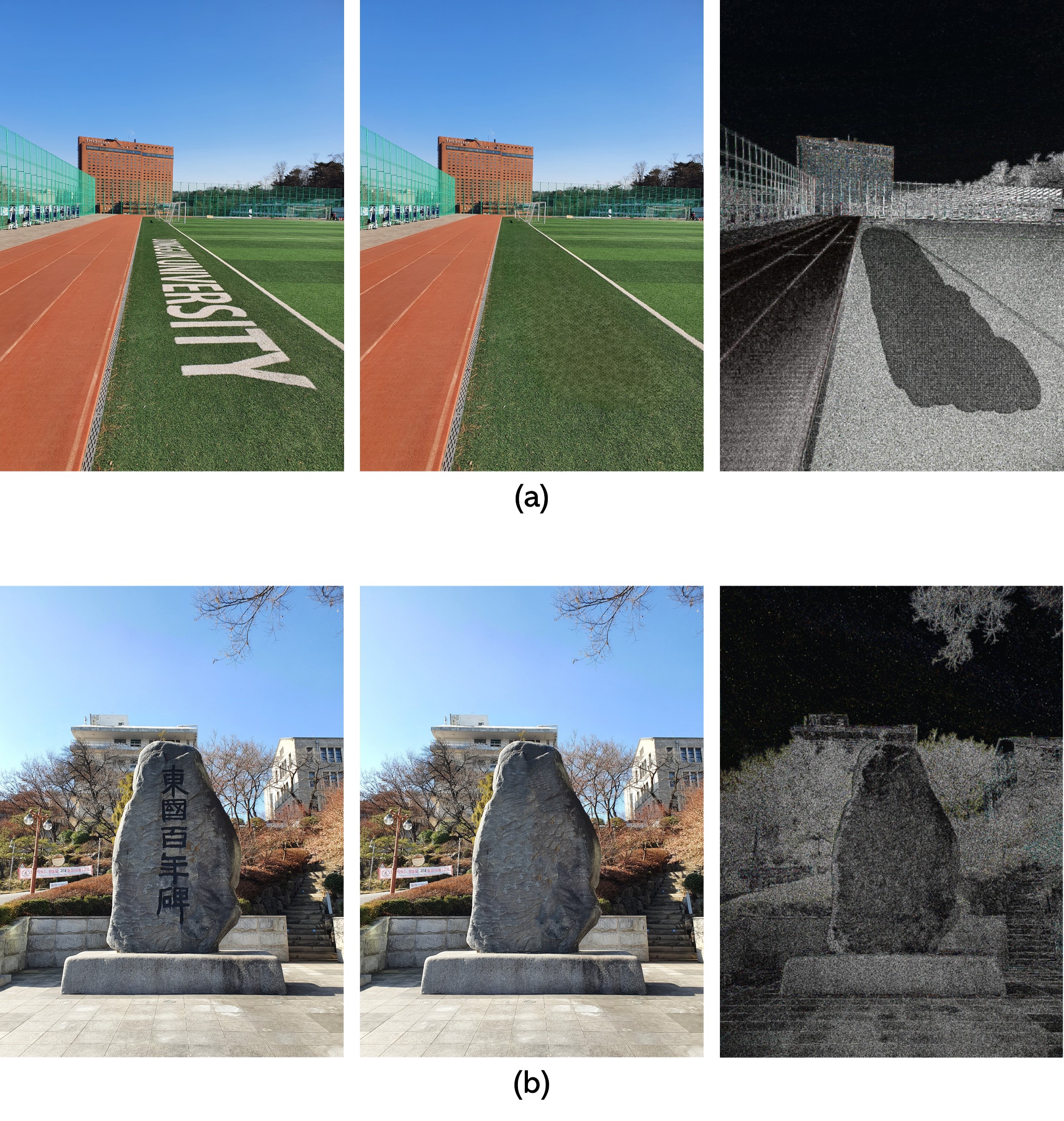}
    \caption{Examples of object removal and noise analysis}
    \label{fig:discussion_example3}
\end{figure}

The application of image processing techniques enables robust analysis by examining the graphic data of the image itself, allowing for resilient analysis against changes in the metadata within the file \cite{vadrevu2022image}. It also possesses the advantage of being able to detect manipulated regions. Figure \ref{fig:discussion_example3} shows the results of detecting manipulated regions using an automated tool that performs noise analysis, generating distinct noise patterns in manipulated regions through the use of a separable Median Filter \cite{chen2022snis, liang2021convolutional, wagner2020forensically}. In Figure \ref{fig:discussion_example3}, (a) shows an example of detecting a manipulated region. However, this technique inherently carries both false positives and false negatives. With the advancement of sophisticated image editing techniques driven by AI, detecting manipulation through existing image processing has become more challenging. In Figure \ref{fig:discussion_example3}, (b) shows the limitations of such image processing techniques. Moreover, the fact that such sophisticated image editing techniques are inherently provided in smartphone applications highlights their disadvantages, as manipulated images can be easily generated.

Mobile forensic artifacts analysis allows for determining when image manipulation occurred and identifying the manipulated regions. In some cases, it is possible to acquire the original images of manipulated images, providing specific information related to the manipulation. However, this technique has the disadvantage of requiring the acquisition of the smartphone used to edit the target image and can only be applied if the relevant app has not been uninstalled.

We have integrated each technique into a unified methodology to mutually complement their disadvantages. For instance, even if the metadata is compromised, image processing (refer to Figure \ref{fig:discussion_example3}(a)) or mobile forensic artifacts can be utilized to determine the presence of image manipulation and identify the manipulated regions. Alternatively, in cases where sophisticated image manipulation techniques are applied and not detected through image processing (refer to Figure \ref{fig:discussion_example3}(b)), manipulation can be detected at other stages proposed in our methodology.

\section{Conclusion}
\label{sec:conclusion} 
This study provides a comprehensive and systematic approach for image manipulation detection through image file analysis, image processing techniques application, and mobile forensic artifacts analysis.

Through the process of image file analysis, we were able to detect image manipulation by examining parsed Exif, DQT, and Filename Signature with the Reference DB. Furthermore, conducting analysis on mobile forensic artifacts enabled us to identify artifacts related to image editing tools. In this study, we considered specific information for accurate image manipulation detection, including the presence of edited images, the ability to identify manipulated regions, the verification of original images used in editing, the existence of edit logs, and the image caching status.

Most image editing tools exhibited artifacts that are relevant to image manipulation, demonstrating their significant utility in the field of digital forensic investigation and analysis. Moreover, by combining image file analysis, which relies on metadata, with image processing techniques that rely on graphic characteristics, we can notice a reduction in false positives in image manipulation detection. The integration of these approaches appears to be effective in improving the overall accuracy and reliability of image manipulation detection.

Furthermore, we shared a dataset of manipulated images. We believe that this dataset will be valuable for research papers analyzing graphical differences in post-processing stages across various applications or focusing on detecting manipulated regions. However, it is important to note that this dataset is based on 11 original images from which the same object or background was removed, and thus, separate masks for the manipulated images are not provided. Additionally, to facilitate image manipulation detection, we provide a code that automates the parsing, analysis, and insertion of manipulation-related metadata into the Reference DB, which serves as the criterion for detecting of image manipulation. This database and code are shared to enable anyone to utilize them for the purpose of image manipulation detection. By sharing the dataset and providing the automation code alongside the Reference DB, we aim to make a contribution to the broader research community and foster research on image manipulation detection.

As future research, we plan to expand the types of image file formats and conduct studies to identify and analyze mobile device's system artifacts that can help identify meaningful user behavior. Through this, we aim to develop more accurate image manipulation detection techniques and procedures, which will result in valuable information for digital forensic investigation and analysis.

\bibliographystyle{elsarticle-num}
\bibliography{sample.bib}

\end{document}